\magnification=\magstep1  
\vsize=8.5truein
\hsize=6.3truein
\baselineskip=18truept
\parskip=4truept
\vskip 18pt
\def\today{\ifcase\month\or January\or February\or
March\or April\or May\or June\or July\or
August\or September\or October\or November\or
December\fi
\space\number\day, \number\year}

\centerline{\bf Area constrained  SOS Models of Interfaces.   }
\medskip
\centerline{by }
\centerline{J. Stecki }
\vskip 20pt
\centerline{ Department III, Institute of Physical Chemistry,}
\centerline{ Polish Academy of Sciences, }
\centerline{ ul. Kasprzaka 44/52, 01-224 Warszawa, Poland}
\bigskip
\centerline{\today} 
\vskip 60pt

\centerline{\bf{ Abstract}}
The solid-on solid (SOS) model in two dimensions ($d=2$)
is now solved  under the constraint of constant 
energy and then under the new constraint of constant total area.
From the  combinatorial factors $g(E;L,M)$, the new ensemble is constructed
with its free energy $F(A_{tot},T)$ of a membrane of 
constant  (onedimensional) area  $A_{tot}$.
 The entropy per column
$Y=(1/L)\log g(E;L,M)$ of rectangular 
$L\times M$ strips reduces to a common curve in reduced variables.
Definitions of the "area" 
and of the interfacial tension, are compared or discussed.
  Analytical calculations are
supported with numerical ones and vice versa.
Overall the constraint reduces the ratio of $A_{tot}$ to the projected
area $L$, as compared with canonical calculation, strongly at high 
temperatures.
 
\vfill\eject

{\bf I. Introduction}

The solid-on-solid (SOS) model originated very long time ago in
connection with crystal surfaces and crystal growth; its first extensive
description is found in the classical paper of Burton, Cabrera, and Frank[1].
 It was later realized that the model of irregularities of a crystal face
 could be fashioned into a theory of morphology and also of fluctuations
 of surfaces. Then it was discovered that the SOS model of equilibrium 
 surface could be solved exactly in two dimensions, {\it i.e.} for a
one-dimensional interface, by the method of transfer matrix [2-7].
 There are reviews available[8-10].
The solution could be applied to free energy calculations and to the theory
of capillary waves[7,11,12] of interfaces.
 
Here we apply the SOS model to the difficult problem of membrane fluctuations.
All interfaces and  all membranes as well, undergo shape fluctuations. These
necessarily result from thermal motion and are intertwined with all 
thermal fluctuations.
 However, there are important distinctions between interfaces and membranes.  
 An interface is an open system, also open with
respect  to particle exchange. A flat liquid surface  increases its surface  
area and takes a wavy form by inducing the diffusion of molecules  
from either bulk phase. Similarly a crystal surface is formed and reformed
by processes of evaporation - by particle exchange with the surroundings.

In fluctuating membranes such diffusion does not occur or is very rare. 
Commonly membranes and bilayers are formed by self-assembling 
surfactant (am\-phi\-phi\-lic) molecules and embedded in a liquid solvent. 
Although in principle the mo\-le\-cu\-les forming 
a bilayer are soluble to some extent in the liquid solvent, 
in reality their concentration is exceedingly low. It is hardly
possible for a bilayer or a membrane to absorb or release molecules in 
the process of locally changing its shape; consequently the shape 
fluctuations (also long-wavelength fluctuations, such as capillary 
waves) take 
place {\it under the constraint} of a constant particle number $N_s$ forming 
the sheet of surfactant.
 This constraint translates to a constraint of constant membrane area. 



Such considerations will make distinction between the true area and the 
"projected area". The exact and formal results provided so far by the Gibbs 
statistical mechanics, always refer to the area defined in terms of the
geometry of the macroscopic volume $V$ enclosing the system under 
consideration. For example in d=3 the volume may be a parallelepiped and 
the area its $x-y$ edge. A constant-volume  increment $\delta F$ of 
the free energy $F$ provides the interfacial tension $\gamma$ 
through 
$$ \delta F =\gamma \delta A~.  \eqno(1.1)$$
$A$ is often called the "projected area", to emphasize  
that, because of the undulations of the interface, its "true area" is 
larger and, when projected onto the edge of the box, produces $A$.
 The "true interface area" is something the external world does not 
have control of; contrariwise, the projected area $A$ is under 
the control of 
external forces just as the volume, temperature, or overall density is.  
Thus (1.1) defines $\gamma$ in terms of a measurable quantity $A$.
Such reasoning clearly applies to an interface; we examine    
the other case of a membrane with a constant intrinsic  area.

   These observations prompted us to consider a new set of SOS models
with constraints. In this paper we limit ourselves to one-dimensional
string embedded in two dimensions, {\it i.e.} to a model which is exactly
soluble[2-7] albeit without constaints.

   We solve below (for the first time) 
the microcanonical {\it i.e. } constant-energy version 
of the SOS model which we use to obtain the combinatorial factors;
these are used to construct the new ensemble of  constant interface 
length.  Therefore we proceed via exact enumeration.

 In Section II
the SOS model is defined and solved; this provides the 
combinatorial factor $g(E;L,M)$ -  the
basic building block for subsequent calculations. 

These results/calculations have many ramifications, some of
which are reported in Section II and in the Appendix 
because of their novelty and possible usefulness. 

 In Section III 
 the definitions of the total area are compared and 
 then  
 the relation between total area $A_{tot}$ and the projected area
 $L$ is compared for the unconstrained calculation and for the 
 new calculations under the constraint of a given constant $A_{tot}$.
 Also, the 
derivative of the free energy with respect to either area 
(defining the tension $\gamma$) is calculated and discussed.

Finally Section IV is the summary and brief discussion.

\bigskip
{\bf II. The Solid-On-Solid Model and its Solution. }

In the two-dimensional Solid-on-Solid (SOS) model[1-9]  
a one-dimensional interface is drawn along a strip of a square lattice 
as illustrated in Fig.1. The interface is
reduced to a one-dimensional string. 
 The standard $L\times M$ strip is  made
of $L$ columns of height $M$; the interface crosses each column 
once and only once.
A microscopic configuration of the interface is then specified by a 
 collection of heights
$$ \{h_j\}, j=1,L ;~~ 1\le h_j \le M   .            \eqno(2.1)$$
 Clusters and overhangs are absent.
The "vertical" variable is $h$ or $z$ at the horizontal position 
$x=a_0 j$, $1\le j\le L$ and we take the lattice constant $a_0=1$. 
The "height" is limited to  $1\le h\le M$. 

In the standard  SOS model[2,5,6]  the energy   of
 a microscopic configuration is 
 $$E_{tot}=E_0 +E  \eqno(2.2)$$ 
with 
$$ E_0 =\epsilon_0  L;~~~ E=\epsilon  A_v.  \eqno(2.3)$$
In general $\epsilon_0 \ne \epsilon$ and both are positive.
$E_0$ is the energy of a perfectly flat interface.

The total length of the interface  is its one-dimensional area;
it is a sum of its horizontal part, $A_h$ and its vertical part, $A_v$.
$$ A_{tot} = A_h +A_v = L + A_v       \eqno(2.4)$$
The first part is the trivial contribution equal to $L$. The vertical 
part is the sum of absolute values of the height jumps 
$$ A_v = \sum_i\Delta_{i,i+1} = \sum_i \vert h_{i+1} - h_i \vert  \eqno(2.5)$$
where
$$  \Delta_{i,i+1}\equiv \vert h_{i+1} - h_i \vert ~,~ \eqno(2.6)$$

Thus the energy of the strip and of the interface is directly
related to the vertical part of the interface area, $A_v$.
This is not essential.
 
 In more general SOS models where
$$ E = \epsilon \sum \vert h_{i+1} - h_i \vert ^n~~~~\epsilon >0~ \eqno(2.7)$$
e.g. when $n=2$ in the Gaussian model,  the simple proportionality of 
$E$ and $A_v$, is lost. 

Also, other definitions of the total area, can be given (see Section III) but
are numerically close to (2.4).

We define the following partition functions and their associated
thermodynamic potentials.
For the finite strip $L$ columns long and $M$ rows high, we  have
{\item {(A)} the microcanonical constant-energy ensemble }
$$ Z(L,M,E) = g(A_v;L,M) ~~~~~~ S=k\log Z   \eqno(2.8)$$
where $g$ is the number of microscopic configurations and $S$ is the 
entropy.  $k$ is the Boltzmann constant, $E=\epsilon A_v$.
Given $\epsilon_0$, one can use interchangeably $E_0$ or $L$.
Given $\epsilon$, one can use interchangeably $E$ or $A_v$.
{\item {(B)} The canonical (constant temperature) partition function is}
$$ Z(T,L,M) =\sum_{A_v\ge 0} g(A_v;L,M)\times \exp^{-\beta E_{tot} }\eqno(2.9)$$
$$ \beta F = (-)\log Z \eqno(2.10)$$
where $F$ is the free energy. This  usually is written subtracting 
the energy of the flat  interface $\epsilon_0 L$ as the trivial term, defining
$$ Z(T,L,M) =\sum_{A_v\ge 0} g(A_v;L,M)\times \exp^{-\beta E }   \eqno(2.11)$$
so that $F=-kT\log Z$ is now the free energy of interface deformations.
By introducing the quantity  $Q = \exp[-\beta\epsilon]\in [0,1] $
we rewrite (2.11) as 
$$ Z(T,L,M) =\sum_{A_v\ge 0} g(A_v)\times Q^{A_v}   \eqno(2.12)$$
emphasizing that $Z$ is the generating function for combinatorial 
factors $g$.
Here, $L$ being fixed, the total length of the interface $A_{tot}$ 
is directly given by $A_v$ and conversely.  
 $E$, $A_v$, and $A_{tot}$ fluctuate - appropriately for an interface.

Generalizing the partition function for fixed $L$, we construct now

{\item{(C)} the new partition function for total interface area.}
That is, we consider an ensemble of strips with different lengths $L$
(and common height $M$). Then
$$ Z^\ast(T,A_{tot};M)=\sum_L \sum_{A_v} g(A_v;L,M)\exp[-\beta E_{tot}]
\eqno(2.13)   $$
with the implied constraint of a given total area, e.g. defined by (2.4)
(as $A_{tot}= L + A_v $). 
Total energy  as given by (2.2)-(2.3) 
takes into account the energy cost of a horizontal step,
$\epsilon_0$, besides the energy cost of a vertical step, 
$\epsilon$. It is convenient to define
$$Q = \exp[-\beta\epsilon]~~~R = \exp[-\beta\epsilon_0],   \eqno(2.14)$$ 
whereby 
  $\exp (-\beta E_{tot})$ is written as $Q^{A_v}R^L$. 
The partition function (2.13) becomes $Z^\ast(Q,R,A_{tot};M)$.
By allowing different energies for vertical and horizontal steps we can 
distinguish and treat as different, total lengths that are equal but are made of 
different proportions of vertical/horizontal steps. 
Both $Q\in [0,1]$ and $R\in [0,1]$. Thus in (2.12) $R=1$ has been
tacitly assumed. 
 The height $M$  can be made infinite[7,8] for fixed $L$ (Section IId).
Under certain conditions, infinite $M$ can also be introduced into (2.13).
The averages are calculated as usual: 
$<L>=-(\partial \log Z)/(\partial \log R)$ and 
$<A_v>=-(\partial \log Z)/(\partial \log Q)$.

To summarize, we consider the following geometries:
{\item{(i)}  L fixed, M fixed and finite,}
{\item{(ii)}  L fixed, M infinite,}
{\item{(iii)}  L variable, M fixed and finite,}
{\item{(iv)}  L variable, M infinite.}\hfill\break
\noindent 
The new partition function for (iii) or (iv)
with the constraint of given constant total area  $A_{tot} = L + A_v$, 
is calculated in Section IId. 

In all cases we need the 
basic building blocks: the combinatorial factors $g(A_v;L,M)$ which are
computed in Section IIa,IIb, and $g(A_v;L,M=\infty)$ in Section IId.

\bigskip
{\bf IIa. The Microcanonical Ensemble of Fixed Energy.}

To solve the SOS model in $d=2$ for both $M,L$ finite and given, 
one introduces the 
$M\times M$ transfer matrix as follows
$$ T(h,h') = Q^{\vert h-h' \vert}     \eqno(2.15)$$
with $Q\equiv \exp[-\epsilon /kT] $.  The canonical
partition function  is expressed as[2]
$$ Z = {\rm Tr}~ T^L = \sum_h (T^L)_{h,h}~.  \eqno(2.16)$$
If the 
transfer matrix is diagonalized, as a real symmetric positive definite 
matrix it admits a representation
$$ T_{hk}= \sum_j \phi_j(h)\lambda_j \phi_j(k)  \eqno(2.17)$$
where $\lambda_j$ is the j-th eigenvalue, $j=1,...,M$ and $\phi_j$ the 
corresponding  eigenvector of the orthonormal and complete set.
Therefore multiplication  of $T$'s just raises the power(s) of eigenvalues
and 
$$ Z = \lambda_1^L+ \lambda_2^L+\cdots~~   .     \eqno(2.18)$$
 The largest eigenvalue 
 dominates in the limit $L\rightarrow\infty$; 
then for sufficiently large $M$,~ 
 $F/(LkT)=-(1/L)\log Z = -\log \lambda_1 = -\log(1+Q) +\log(1-Q)$ 
reproduces the known result,  $\log \tanh \beta\epsilon/2$ [2,7,10,12]. 

Now we want to select from the set of all configurations only those 
of given total length $L+A_v$.
If $L$ is fixed this is reduced to a constraint of given $A_v$. The latter is 
$$ A_v= \Delta_{12}+ \Delta_{23}+ \cdots+ \Delta_{L,1} ~,~\eqno(2.19)$$ 
where  
$$  \Delta_{i,i+1}\equiv \vert h_{i+1} - h_i \vert ~,~ \eqno(2.6)$$

The last term in (2.19) is 
there if one imposes periodic boundary condition $h_{L+1}=h_1$ as we do. 
The constraint on $A_v$ can be expressed with the aid of a Kronecker delta 
$\delta^{Kr}(j,n)$ equal to 1 for $j=n$ and to 0 otherwise. 
Thus
$$ Z(T,L,M;A_v) = \sum_{h_1}...\sum_{h_L}\delta^{Kr}(A_v,\Delta_{12}+...) \times
T_{h_1,h_2}~ T~ T \cdots~~~        .    \eqno(2.20)$$
By using the integral representation
$$ \delta^{Kr}(j,n)=\int_{-\pi}^{+\pi} ({{dk}\over{2\pi}}) e^{ik(j-n)} 
\eqno(2.21)$$ 
and taking the integral in front of the sums, we find that to each 
T-matrix there corresponds a factor $\exp(ik\Delta)$ so defining
the new matrix
$$ \tau_{h,h'} \equiv T_{h,h'}e^{ik\Delta_{h,h'} }~~~, \eqno(2.22)$$
we have the new expression for the constrained partition function
$$ Z(T,L,M;A_v) =\int_{-\pi}^{+\pi} ({{dk}\over{2\pi}}) e^{-ikA_v} {\rm Tr}~\tau^L~~~
.   \eqno(2.23)$$
This is the expression we evaluate. The matrix {\bf $\tau$ } 
can be put into the form of the r.h.s. of (2.15);
define $\tilde Q$ as $\tilde Q\equiv Q\exp(ik)$,
so that
$$ \tau(h,h')=\tilde Q^{\vert h-h' \vert}~~~~ \tilde Q\equiv e^{+ik-\beta\epsilon}~~ .   \eqno(2.24)$$
The complex $\tilde Q$ has a real part between -1 and 1 and an imaginary part
between -1 and 1.
The  matrix {\bf $\tau$ } depends parametrically on the wave-vector $k$, the
integration variable, which is real. It reduces to the original T-matrix at $k=0$, 
apparently continuously. By using the package "Eispack"[13], in particular
its program CG for diagonalization of complex  general matrices, we were
able to compute $Z$ and log$Z$ from
$$ Z(T,L,M;A_v) = \int_{-\pi}^{+\pi} ({{dk}\over{2\pi}}) e^{-ikA_v} [
\lambda_1(k)^L+\lambda_2(k)^L+\cdots +\lambda_M(k)^L]~ .  \eqno(2.25)$$
The eigenvalues depend on $k$. Evaluation was done for $M=5,6,10,20$
and for $L < 41$; several values of $Q$ were used. It could be done for $M=30$. 
For odd $A_v$ the result is exactly nil and this is correct because of the
periodic boundary condition at $L$. For $A_v=0$, $Z=M$ for all $Q$. 
The general form of the result is
$$Z(A_v)=g(A_v)Q^{A_v}~ .  \eqno(2.26)$$
with $g$ independent of $Q$. The Kronecker delta picks up the one 
term of the sum (2.12). By dividing (2.26) by $Q^{A_v}$ we obtain $g(A_v;L,M)$.
All numerical results did conform to (2.26); $g$ was obtained accurately  
for the very largest systems ($g$ must be an integer)  and was 
obtained exactly for all other sizes. Numerically, $Q$ cannot be chosen
too small or too close to unity.
In actual calculations we have imposed periodic boundary condition 
$h_{L+1}=h_1$ which not only simplified the calculation but also excluded
exotic configurations. For all $A_v$ odd, $g$ is nil. 

The combinatorial factor $g(A_v)$ always follows the same
pattern starting from $g(0)=M$, going through a maximum near $A_v\sim LM/3$ 
and then falling down to 0 for $A_v >(M-1)L$. Being the number of microscopic 
configurations, $g$ very soon reaches horrendously large numbers and for
that reason Fig.2 shows an example of $g(A_v)$ for very small $M,~ L$.
The pattern seen in Fig.2 is general; starting with $g(0)=M$ and after a
very rapid increase, $g$ goes through  a maximum and falls to a low value,
often terminating with a value of~~2. Afterwards for all values of $A_v$ 
the correct value is $g=0$. 
This characteristic dependence becomes  continuous for large $L$ 
(and not too small $M$ ) in  the following variables: $ Y=(1/L)\log (g)$ 
and  $a\equiv A_v/Ln_1$. The quantity $a\equiv A_v/Ln_1$ varies between 
0  and 1~~, $n_1$ being equal to $M-1$.

Since the energy is not allowed to fluctuate, the logarithm of the
partition function is  the  thermodynamic potential of the
constant-energy  microcanonical ensemble 
$$ S(L,M,E) = k \log g(A_v;L,M)  ~~~ E=\epsilon A_v    \eqno(2.27)$$

An extension of the numerical results to larger systems, can be based 
on a simple scaling in reduced variables. 
We have already introduced the scaled, or reduced area (simply area per site) 
$a\equiv A_v/L(M-1) $  and we find that $Y=(1/L)\log(g)$ fall on a common 
curve when plotted against $a$, for different $L$ at each constant $M$.
  
 Fig.3 shows how this simple "scaling" works; the shapes are similar enough so that
  for each M  different values of L fall on a common curve.
 Values  of $L<10$ and of $M<10$ are too small to be included.

Eq.(2.27) identifies the function $Y$ as entropy per column. Invoking the
 relation 
$$ dS/dE = 1/T \eqno(2.28)$$
defining the thermodynamic temperature $T$ in the microcanonical ensemble,
we may rewrite it as $d\log(g)/dA_v = \beta\epsilon = -\log Q$ with $\beta = 1/kT$.
However, as a look at Fig.2 and Fig.3 shows, beyond the maximum in $g$
that interpretation fails (or produces negative temperatures). The 
original observation is attributed to L.Landau[14]: 
in systems with energy bounded from above, $T<0$ is possible.
Here is one more such example.

The ordinary canonical partition function is calculated by 
diagonalizing the real $T-$matrix for given $Q$, {\it i.e.} after (2.16-18).
But it is also equal to (2.12) . Since all terms are positive, 
selecting the maximum term leads to $d(\log g)/dA_v -\beta\epsilon =0$
which can be rewritten to a form like (2.28). Now $T$ is an independent
variable, $T>0$ because $Q\in [0,1]$, and $E=E(T), S=S(T)$.  

\bigskip
{\bf IIb. An Independent Algebraic Check.}

Turning to the starting point for the solution of the model, (2.15-2.16),
 we note that for  small $L$ the trace can be computed analytically.
 By multiplying out the $T$-matrices, we obtain a polynomial in $Q$:
$$ Z=Tr T^L = c_0 + c_1Q + c_2Q^2+... ~~ . \eqno(2.29)$$
The basic observation is that the power of $ Q$ is just the number of 
energy units,  identified in the previous subsection as the vertical part 
of the total length and  denoted by $A_v$. $Z$ containing the total  of 
all configurations, each term represents the split between different
values of $A_v$. That is , we can identify the coefficients $c_j$ in 
(2.29) as the combinatorial factors denoted  by $g(A_v)$.
 Thus equation (2.29) is rewritten as
$$ Z = \sum_{A_v=0,1,2...} g(A_v)Q^{A_v}~.  \eqno(2.12)$$
For example, constructing the 4 by 4 $T$-matrix (whose first row is
$ 1,Q,Q^2,Q^3$)  and multiplying it out to obtain $T^6$, taking the
trace, we obtain for $L=6,M=4$~~	 
$Z=4+90x^2 +510x^4 +1266x^6 +1116x^8 +744x^{10} +310x^{12}
 +42x^{14} +12x^{16} +2x^{18} . $
Thus $g(A_v=8) =1116$ etc. See Fig.2. Such exact enumerations, with the
 computer-aided algebra, were carried out up to $M=20$ 
for  $L$ up to 30, providing checks on numerical results obtained via the 
diagonalization  described in previous Subsections. 

\bigskip
{\bf IIc. The Constraint of Total Area For Given M}\hfill\break

From the combinatorial factors in the microcanonical ensemble at given $L,M$ 
we can construct now the partition function with the constraint of total length 
$A_{tot}$. We take the ensemble of strips with different $L$'s (and the same
$M$) and implement the constraint of constant $A_{tot}$ in (2.13);
for given $M$ 
$$ Z^\ast(T,A_{tot};M)=\sum_L \sum_{A_v}\delta^{Kr}(L+A_v, A_{tot}) 
g(A_v;L,M) Q^{A_v} R^L.    \eqno(2.30)$$
That is, the sum $L+A_v$ is  held constant. 
 Maximum value of $L$ is obviously $A_{tot}$
and the mimimum value is 2. To avoid exotic cases, periodic boundary 
conditions are imposed $h_L=h_1$. The Kronecker delta is replaced
by these simple restrictions on the sums.

Note that here we do not calculate the average total area
$A_{tot}$ because the procedure
is inverted: for {\it given} fixed total length $A_{tot}$ both $L$ and $A_v$ 
are allowed to fluctuate and are calculated as averages: 
$$ \langle L\rangle=d\log Z^\ast/d\log R~~~\langle A_v\rangle=d\log Z^\ast/d\log Q \eqno(2.31)$$
and $\langle A_v\rangle=A_{tot} -\langle L\rangle $ or conversely. 
The averages as $f(Q,R,A_{tot})$ were 
computed and are shown below for special cases 
such as $u\equiv Q=R$ and $R=1, Q\in [0,1]$. 
In the first 
case a string of fixed length is allowed to fluctuate in a neutral
environment, where energy of a horizontal step is equal to that of a
vertical step and therefore none is preferred; 
in the second case, only the energy of vertical steps is 
affecting the choice of configuration.  

Interestingly, the conbinatorial factors $g(A_{tot})$ are increasing without 
bounds, in contrast to the constant $L,M$ case from Section IIa-IIb.

  
For an ensemble of narrow strips of $M\ge 6$
analytical calculations were done as follows: first traces 
of $T^L$ were calculated and stored
(for $L$ up to 40 and nore) and then sums were formed for all even values of 
$A_{tot}\in [10,38]$. From the resulting expression 
$\beta F, \langle L\rangle, \langle A_v\rangle$ were calculated 
analytically and used, in part
to check the results of numerical computations. 
A selection of these results is shown in Section III. 
The results for finite $M$ are also
compared with the case $M=\infty$ described below.

\bigskip
{\bf IId. The Columnar Strip (M=$\infty$) and its Solution}.\hfill\break


From the solution[7] of the eigenvalue problem of $T$ for given $L$ 
and columns of infinite height[7], $M=\infty$,  the eigenvalues 
are known explicitely 
$$   \lambda (\nu) = (1-Q^2)/(1+Q^2 -2Q\cos (\nu) ) \eqno(2.32)$$
where the index $\nu\in [-\pi,+\pi]$ is now continuous. 
Because of the translational invariance in the "vertical" $M$ 
 direction, actually exploited in solving the transfer matrix[7],
we must impose a constant $h_1$, e.g. $h_1=0$ and another fixed value 
for $h_L$ - which we 
choose to be equal to $h_1$. The canonical partition function (2.11) or
(2.12) with $M$ infinite then exists. There are some limitations 
in more general cases ($R\neq 1$ and variable $L$ fixed $A_{tot}$) 
- because now the excursions of the interface are limited only by 
the cost of a vertical step.  The 
flat configuration $h_i=0$  is the equilibrium configuration
for $Q\rightarrow 0$.

We need the combinatorial factors $g(A_v;L,M=\infty)$. Following 
 Section IIa, in order to impose the 
 the constraint of fixed $A_v$ we
replace $Q=\exp(-\beta\epsilon)$ with $\tilde Q \equiv Q e^{ik} $.
The solution[7] goes through as explicit calculation shows.
The partition function becomes a double integral, replacing (2.25) by
$$ Z(Q,A_v;L,M=\infty) = \int_{-\pi}^{+\pi} ({{dk}\over{2\pi}}) e^{-ikA_v} 
\int_{-\pi}^{+\pi} ({{d\nu}\over{2\pi}})\lambda (\nu)^L~ .  \eqno(2.33)$$
Numerical evaluation  is now simpler as the diagonalization step is 
not needed  -  we have the explicit expression (2.32) for the eigenvalues. 
For $A_v$ and $L$ integers, we obtain again $Z$ in the form (2.26) {\it i.e.}
$ Z=g\times Q^{A_v}  $,
where again $g(A_v)$ comes out an integer. Again $\log g$ has the 
interpretation of entropy $S/k$. 

In an analytic calculation, all $g(A_v;L)$ result as polynomials in $A_v$.
There is a striking simplification which occurs for $M=\infty$;
 $g(A_v;L)$ are
polynomials with highest power equal to $L-2$. For example
$g(x;L=4)=2+(2/5)x^2$ (for all $x>0$), $g(x;L=6)=2 +(35/8)x^2 +(21/32)x^4$ 
etc.~~($x\equiv A_v$).
Knowing these polynomials we can make computations with {\it any value} of
$A_v$. A few examples are given in the Appendix.

The exact results for $g(A_v;L,M=\infty)$ are used now to construct  an
 ensemble of columnar strips of infinite heights $M$ and variable 
 $L$. At given $A_{tot}$, 
 maximum value of $L$ is  $A_{tot}$,
and the mimimum value is 2. Minimum value of $A_v$ is zero; there is no
upper bound other than $A_{tot}-L$.  
To avoid exotic cases and to ensure the existence of $Z$, 
again $h_L=h_1=0$ for each $L$. 
We show  here,  in Section III, explicit calculations for these two cases: $Q=R\rightarrow u$
and $R=1, Q\in [0,1]$.

\bigskip
{\bf III. Numerical Results and the Comparison of Areas and Free Energies.}

In this section we show a restricted selection of results for the two 
most important quantities: the vertical area $A_v$ related to the energy and 
the total area of the interface.
 We also consider  the derivative of the free energy $F$ with respect
 to area, all other variables being kept constant; this is interpreted as
 the interfacial tension $\gamma$.
 Hence
 $$ \beta \gamma = (\partial (\beta F) /\partial A)_T \eqno(3.1)$$
The area $A$ can be the projected area $L$, the total area imposed in the
ensemble of all $L$'s, or the average total area defined as $\langle A_{tot}\rangle
\equiv L+\langle A_v\rangle $. All results are given  for $R=1,Q\in
[0,1]$  ~(cf.(2.13)-(2.15)).
The other interesting case, $R=Q\in [0,1]$, turns out to be a degenerate
temperature-independent one, in which the interface wiggles randomly in 
an energetically neutral environment. These results are also reached in the
limit $Q\rightarrow 1^-$ at constant $R=1$.

First, however, we compare 3 different definitions of the area, or, rather,
two definitions, $A_v$ and $A_{sq}$, and one approximation, $A'$. Our 
definition of the total area as $A_{tot}= L+A_v$ corresponds to the 
length of the thick line in Fig.1; $L$ is the sum of "horizontal" steps, 
and $A_v$ is the
sum of "vertical" steps, $\Delta_{i,i+1}$ (cf. (2.5-6).
The usual definition would 
draw a line through $h_i$ and $h_{i+1}$; this approach produces
$$A_{sq}\equiv \sum[-1+\sqrt{1+\Delta_{i,i+1} } ]. \eqno(3.2)  $$ 
where $L$ is subtracted.
Expansion of the square root gives the first term, which is the common
approximation, 
$$A' \equiv \sum (1/2)\Delta_{i,i+1}^2 . \eqno(3.3)$$
 Definitions $A_v$ and $A_{sq}$ produce always very close results, $A_v$
always slightly larger.  Fig.4 shows all 
three areas calculated in the canonical ensemble, plotted against $Q$.
$A'$ is a very bad approximation, grossly exaggerating $A_{sq}$ except when
both are very small; this happens here at small $Q$. We have found
earlier in another context[15] the same pattern, with the
commonly used approximation for the area, like (3.3), grossly overestimating
the true area, except in the limit $\Delta A \rightarrow 0$. 
  
  Figure 4 also includes the microcanonical results {\it i.e.} the curve of 
$A_v$ as independent variable against
$Q=\exp(-\langle\beta\epsilon\rangle )$ with 
the inverse temperature computed acccording to (2.28) with (2.27); even
for the small value of $M$ the differences are negligible and invisible
on the scale of the plot. This is a useful check;
the agreement between the canonical and microcanonical ensembles can only
become better for larger $L$ and $M$ and it does.
In conclusion, we use $A_v+L$ as a measure of area, include  
canonical averages, and do not quote microcanonical averages.

Fig.5 illustrates the results for the new ensemble of variable $L$ and
constant $A_{tot}$. Three values $A_{tot}=10,20,30$ were chosen.
Since $A_{\rm sq}$ and $A_v$ are so close, only
 $\langle A_v\rangle$ is plotted. 
 The case $M=\infty$ (section IId) is shown with lines; at the infinite
 temperature $\langle A_v\rangle$ reaches 4.2, 9.2, and 14.2 for 
 $A_{tot}=10,20,30$, respectively. Interestingly, squeezing the membrane
 with a small $M=6$ does not change  $\langle A_v\rangle$ much; these
 averages are shown with stars, crosses, and plus signs, and reach 4.0, 8.6,
 and 13.0. Unconstrained averages calculated in the canonical ensemble  
 produce plots of another shape and much larger values at high 
 temperatures;  $\langle A_v\rangle$ for $M=\infty$ diverges.  For 
 fixed  $A_{tot}$ it does not.
 
We choose the inverse temperature as the ordinate for the plot because
of the quasi-linearity at high temperatures of the data with fixed  
$A_{tot}$. As can be seen clearly,
fixing $A_{tot}$ limits the vertical excursions of the interface
considerably and - somewhat suprisingly - in a similar fashion for
small $M=6$ and for $M=\infty$. 
At low temperatures i.e. at vanishing $Q$ and large $\beta\epsilon$ all
calculations produce $\langle A_v\rangle\rightarrow 0$.

At $Q=1$ {\it i.e.} $\beta\epsilon=0$, $\langle L\rangle$ is at 55-57 
percent of $A_{tot}$,
practically in all calculations we have done so far for the new ensemble.
 The limit 
corresponds to the fluctuation of the interface in the neutral environment
where horizontal or vertical steps carry the same cost in energy, now 
vanishing in the limit $Q\rightarrow 1, \beta\epsilon\rightarrow 0$.

 The divergence of the canonical average for
$M=\infty$ (full line in Fig.5) can be understood as follows.
 If $M=\infty$ and $L$ is
fixed whereas $A_v$ and $A_{tot}$ are free to fluctuate and take an 
average value, it is only the 
energy cost expressed in the value of $Q$ which limits the excursions
of the interface and therefore the number of realized configurations
thus the partition function. When $Q$ ceases to operate, the partition 
function and the averages, diverge. This does not happen when a fixed 
$A_{tot}$ is imposed.

All partition functions are converted to free energies and all free 
energies depend on an area as the independent extensive variable.
The definition (3.1) of the interfacial tension $\gamma$, 
is now applied to $\beta F(A_{tot};T)$.
 In Fig.6 the derivative (3.1)
w.r.to $A_{tot}$ is plotted  against $Q$. For $M=\infty$ or $M=6$  
in the ensemble of all $L$, the data points show mild near linear variation
with $Q$. 

For comparison, the canonical  free energy per column 
$\beta F(Q;L,M)/L$  
 from the largest eigenvalue, is $-\log[(1+Q)/(1-Q)]$ and is 
plotted as the dot-dash line. All eigenvalues for $M=6$ produce the 
near-linear full line. 

Whether the derivative (3.1) applied to $F(A_{tot},T)$ can be interpreted
as an "interfacial tension", is perhaps debatable. Still the derivative
can be calculated and compared with other derivatives of the free 
energy with respect to other areas.

\bigskip
{\bf IV. Summary and Discussion.}

As recalled in the introduction membranes keep their area constant,
notwithstanding shape fluctuations. Planar interfaces are also twodimensional
sheets embedded in three dimensions, but are open to particle exchange
with the surroundings and their proper or intrinsic area varies.
 Hence the theory of fluctuating interfaces must
be suitably modified to apply to membranes. We examine this
modification by
choosing an existing simple and  soluble model, 
namely the SOS model in two dimensions, for which exact enumerations were
practical.

 By starting from the available results, we introduced the 
constraint of constant interface area; first simply into the $L\times M$
strip and then constructing a new {\it ensemble of all projected areas}.
 
The partition functions, free energies, and the ensemble averages 
 all  were built on the
microcanonical combinatorial factors $g(A_v;L,M)$. Their calculation 
was essential.
These were obtained by several means described in Section II.
Of the several calculational schemes, the ensemble of all
lengths $L$ at infinite height $M$ under the constraint of a given total 
area $A_{tot}$ 
produced  $\langle A_v\rangle$ and $\langle L\rangle$. The latter has 
an interpretation of the average projected area. Practically in all 
calculations, the ratio  $\langle L\rangle /A_{tot}$ approached 55 percent
in the limit of high temperatures ($Q\rightarrow 1,\beta\epsilon\rightarrow 0$).

We also calculated the derivative of the free energy $F(A_{tot},T)$ 
with respect to $A_{tot}$. This may be considered an analogue of the 
interfacial tension  $\gamma$, now 
defined with respect to  $A_{tot}$. When compared with other calculations
of similar derivatives, the new derivative was the smallest in the absolute 
value and did not diverge in the limit of infinite temperature. This was 
illustrated by Fig.6.

The mathematical procedures involved some particular features; their 
description may be useful.
First, we used a diagonalization of a complex  
matrix ${\bf \tau}$ which was symmetric but not unitary. This was followed 
by numerical integration. The final result had to be an integer.
The correcteness of the diagonalization procedure was verified by 
agreement with the exact algebraic calculations.
 The exact enumerations (for the case of infinite $M$) were also spot-checked 
by the generating 
 functions for "partially directed" random walks, obtained by slightly 
modifying 
 the generating functions given by Privman and Svrakic[18]; but introduction
 of constraints was, as often is the case,  not any more practical.

Extension to two-dimensional membranes embedded in $d=3$ is not impossible though
difficult. A good candidate[17] is the Gaussian SOS model on a square lattice with 
each site endowed with a continuous height variable $h_i$; it is readily 
soluble without constraints and for various boundaries.
The prediction is that the constraint will damp the fluctuations in a 
way similar to the damping seen here.

Finally, it is probably all too obvious that our aim was not to develop further
the Solid-on-Solid model but rather to use it to examine the differences
between membranes and  interfaces. Thus we have not 
pushed the calculations in order to go to highest possible orders of
calculation . The trends are already visible.

\bigskip

{\bf Acknowledgements.}

The author has greatly benefitted in the past from discussions with 
J.S.Rowlinson FRS (Oxford), R.Evans FRS (Bristol), and B. Widom (Cornell).

\bigskip
\medskip

{\bf Appendix.}

{\bf Combinatorial Factors  $g$ for Infinite $M$. }

It is a rare and desirable event if  combinatorial factors can be generated 
by polynomials.
We list the first few combinatorial factors $g(x)$
for $L=$2, 4, 6, 8, 10. These were calculated and are available for $L$ up to 30. 
These polynomials are valid for any value
of $x\equiv A_v>0$. For $x=A_v=0$, $g=1$ for any $L$. 
Some exact algebraic calculations were done in part with the aid of 
an old version of "Mathematica"[16].
See Section IIc for notation and further details.  
\item {L= 2} $g=2$
\item {L= 4} $g=2+(5/2)x^2$
\item {L= 6} $g=2+(35/8)x^2 +(21/32)x^4$
\item {L= 8} $g= 2 + (707/120)x^2 + (77/48)x^4 + (143/1920)x^6$
\item {L=10} $g= 2+(14465x^2)/2016+(4147x^4)/1536+(715x^6)/3072+(2431x^8)/516096$
...

\bigskip

{\bf References.}

\item{1.} W. K. Burton, N. Cabrera, and F. C. Frank, Phil. Trans.   
          Roy. Soc.(London) {\bf 243}, No.866,299-358 (1951);
          Nature {\bf 163},398 (1949). 
\item{2.} J. M. J. van Leeuwen and H. J. Hilhorst, Physica {\bf A107}, 319 (1981).
\item{3.} S. T.Chui and J. D. Weeks, Phys. Rev. B {\bf 23}, 2438 (1981).
\item{4.} T. W. Burkhardt, J. Phys. {\bf A14}, L83 (1981).
\item{5.} D. M. Kroll and  R. Lipowsky, Phys. Rev. B {\bf 28}, 5273 (1983).
\item{6.} A. Ciach, Phys.~Rev.~B {\bf 34}, 1932 (1986).
\item{7.} A. Kooiman, J.M.J. van Leeuwen, R.K.P.Zia, Physica {\bf A170}, 124 (1990).
\item{8.} D. B. Abraham, in {\it "Phase Transition and Critical Phenomena"},
         (C. Domb and J. L. Lebowitz, Eds.), Vol.10, Academic Press, London (1986).
\item{9.} V. Privman, Chapter 1 in  {\it "Finite Size Scaling 
       and Numerical Simulation"}, (Singapore, 1990) where further 
       references can be found. 

\item{10.} J. Dudowicz and J. Stecki, in {\it "Fluid Interfacial phenomena"},
           (C. A. Croxton, Ed.) J. Wiley 1986, p. 637ff.
\item{11.} A. Ciach, J. Dudowicz, and J. Stecki, Phys. Rev. Lett. {\bf 56}, 1482 (1986).

\item{12.} A. Maciolek and J. Stecki, Phys. Rev. B {\bf 54}, 1128 (1996).

\item{13.} Eispack - Fortran package maintained by Argonne Natl. Laboratory,
           version 1983. 
\item{14.}  L.~D. Landau  and L. Lifshitz, {\it "Statistical Physics"} (Pergamon, Oxford 1980).

\item{15.} J. Stecki, J. Chem. Phys. {\bf 114}, 7574(2001), cf.eq.(5.4) ff.

\item{16.} Mathematica - a software package (ver.2.2) by Wolfram Research, Inc. (1994).

\item{17.} J. Stecki, in preparation.

\item{18.} V. Privman and N. M. Svrakic, {\it"Directed Models of Polymers, 
Interfaces, and Clusters"},  Springer-Verlag (Lecture Notes in
Physics No.338), Berlin 1989.

\bigskip
\vfill\eject
{\bf Figure Captions}

{\bf Figure 1}

The solid-on-solid (SOS) interface in $d=2$. The strip is $M$ sites wide 
in the "vertical" direction and 
$L$ sites long in the "horizontal" $x$ direction. The interface is of  
length $A=L+A_v$.

\medskip
{\bf Figure 2}

An example of the number $g$ of configurations of the interface 
plotted against  
the interfacial  area $A_v$ at given size of the strip with L=6,M=4.
The numbers (different from zero) are: 4, 90, 510, 1266, 1116, 744, 
310, 42, 12, 2, 
for areas $A_v$ 0,2,4,6,8,...,18, respectively.

\medskip
{\bf Figure 3}

The "scaling" curve of the microcanonical combinatorial factor $g=g(A_v)$
at fixed $L,M$. The function
$Y=(1/L)\log(g)$ is plotted against $a=A_v/L(M-1)$  for $M=6, 10, 20$, and
$L=10,20,30$. Smaller values of $L$  deviate from
the common  curve. The maxima extrapolated to $1/L\rightarrow 0$
with polynomials in $1/L$ were 1.77702, 2.28587, and 2.97979 for $M=6,10,20$,
respectively. Extrapolation to $1/M \rightarrow 0$ was inconclusive.

\medskip
{\bf Figure 4}

 Three areas: $A_v/L$, $A_{sq}/L$, and $A'/L$ vs. $Q$ calculated 
in the canonical ensemble (see text) are compared. Full line terminating
at 5.8198 - $A_{sq}/L$; bold-dashed line hitting 33.25 - $A'/L$; 
fine-dashed line hitting 6.6500 - $A_v$. Plus signs - microcanonical 
$A_v$ reaching 6.564 at $Q=1$. $L=30$,$M=20$. $A_v/L$ was independent 
of $L$ and weakly depending on $M$ provided $M\ge 10$.

\medskip
{\bf Figure 5}

 Average  $\langle A_v\rangle$ vs. $\beta\epsilon=\epsilon/kT=-\log Q$  
 for fixed $A_{tot}=10,20,30$ in the ensemble of all $L$.  Fixed $M=6$
 shown with points (stars, crosses, plus signs, respectively); $M=\infty$
 shown with lines (fine-dashed, dashed, full, respectively). Unconstrained
 canonical averages $\langle A_v\rangle$ for constant $L,M$ are
 ($M=6, L=10,20,30$, - dotted lines; ditto  
 $M=\infty, L=10$, -  full line) . For $\beta\epsilon=0$,
 data shown with points hit  4.00, 8.564,
 12.980, respectively; data for $M=\infty$ shown with lines, hit 
 4.205, 9.209, and 14.213, respectively. The dotted lines 
(canonical  $\langle A_v\rangle$ for $M=6$) hit 19.444, 38.888, 58.333 
for L=10,20,30,
 respectively. $R=1, Q\in [0,1]$ in all cases. 

\medskip
{\bf Figure 6} 

 Free energy derivatives w.r.to the total area 
$(\partial \beta F /\partial A_{tot})_Q $, plotted against $Q$. 
The new ensemble of all $L$ 
and $M=6$ - small plus signs, ditto $M=\infty$ -large plus signs.
For comparison  the derivatives w.r.to the projected area $L$,
in canonical ensemble $M=6$ - full line, $M,L\rightarrow\infty$ - 
dash-dot line, are plotted.

\bigskip

the end of text. Figures are as 6 *.eps files 

\vfill\eject\end
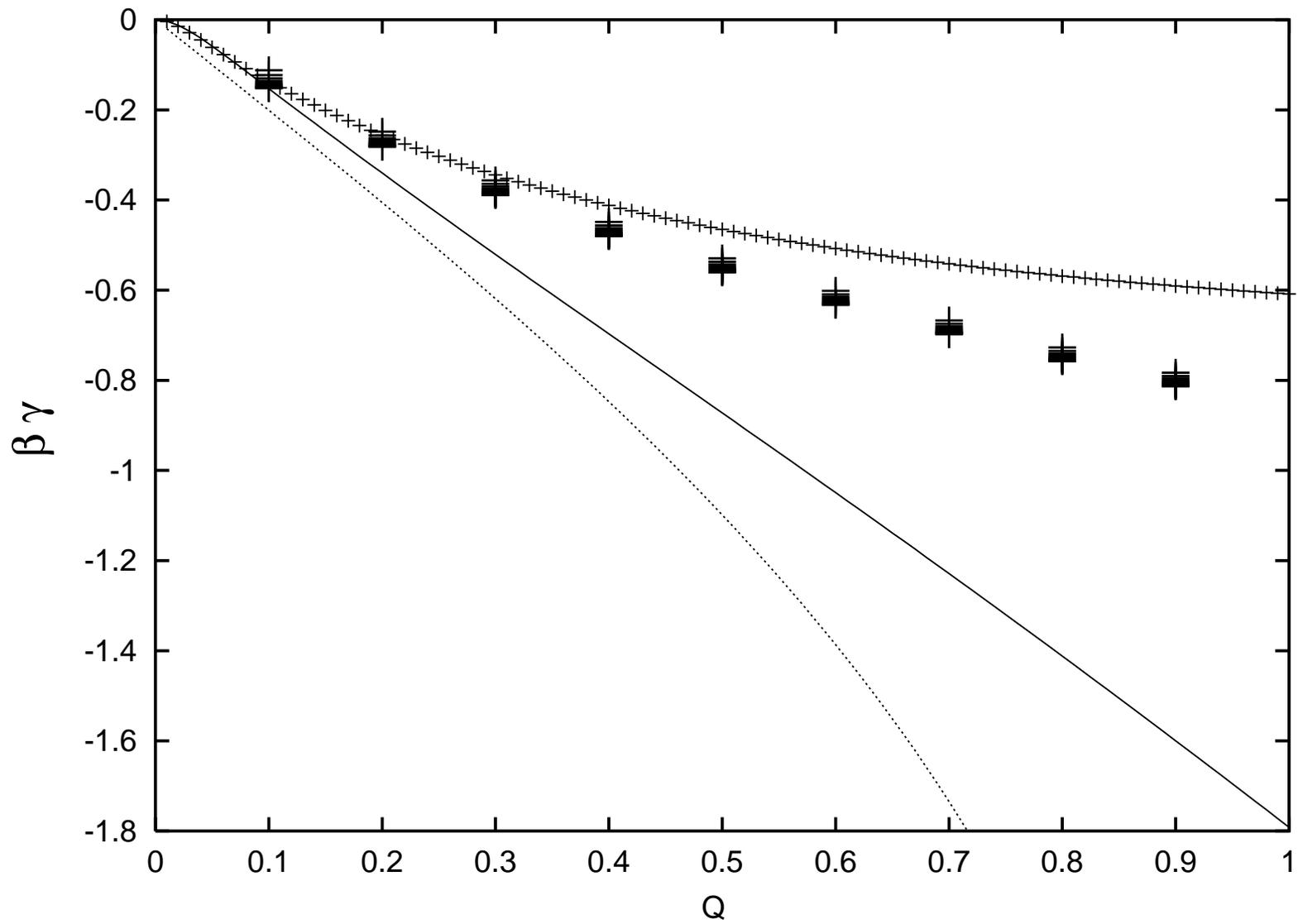
\bye